%% file: inforum2023-fleec-ea.tex
\PassOptionsToPackage{hidelinks}{hyperref}

\documentclass[runningheads]{llncs}

\usepackage{algorithm}
\usepackage{algpseudocode}
\usepackage{graphicx}
\usepackage{subfig}
\usepackage{caption}

\usepackage[utf8]{inputenc}
\usepackage[T1]{fontenc}
\usepackage{zlmtt}
\usepackage{fancyvrb}
\usepackage{booktabs}
\usepackage[T1]{fontenc}
\usepackage{graphicx}
\usepackage{orcidlink}
\usepackage[backend=biber, style=lncs]{biblatex}
\usepackage{xspace}
\usepackage[]{todonotes}

\newcommand{\systemname}{F\kern-.075em L\kern-.2em\raisebox{0.65ex}{ee}\kern-.065emC\xspace}

\newcommand{\CLOCK}{\textsc{clock}\xspace}
\newcommand{\MEMCLOCK}{Mem\CLOCK}

\begin{document}
%
%\title{Non-Blocking Memcached\thanks{Supported by Research Grants …}}
\title{\systemname: a Fast Lock-Free Application Cache}

\author{André J. Costa\inst{1}~\orcidlink{0000-0000-000-000}
\and Nuno M. Preguiça\inst{1}~\orcidlink{0000-0000-000-000}
\and João M. Lourenço\inst{1}~\orcidlink{0000-0002-8495-6442}
}
\authorrunning{A. Costa et al.}
\institute{NOVA University Lisbon — FCT \& NOVA LINCS, Portugal\\
\email{aj.costa@campus.fct.unl.pt} \qquad \email{nuno.preguica@fct.unl.pt} \qquad \email{joao.lourenco@fct.unl.pt}
}
\maketitle              % typeset the header of the contribution
%

\input{extended_abstract}

\printbibliography

\end{document}

%% file: extended_abstract.tex
%Intro
\paragraph{\bfseries Introduction.}
When compared to blocking concurrency, non-blocking concurrency can provide higher performance in parallel shared-memory contexts, especially in high contention scenarios.
This paper proposes \systemname, an application-level cache system based on Memcached~\cite{memcached}, which leverages re-designed data structures and non-blocking (or lock-free) concurrency to improve performance by allowing any number of concurrent writes and reads to its main data structures, even in high-contention scenarios.
We discuss and evaluate its new algorithms, which allow a lock-free eviction policy and lock-free fast lookups.
\systemname can be used as a plug-in replacement for the original Memcached, and its new algorithms and concurrency control strategies result in considerable performance improvements (up to $6\times$).
%Previous:
% In this paper, we present new algorithms that allow the cache to implement an eviction policy in a non-blocking manner, whilst maintaining fast lookups.

%Contributions
\paragraph{\bfseries \systemname Design.}
Memcached~\cite{memcached} has three main separate data structures: a hash table with singly linked-list buckets, for fast lookups; a doubly linked-list to implement a Least Recently Used (LRU) eviction policy; and a slab allocator for fast and efficient memory allocation.
These separate data structures are incompatible with non-blocking concurrency control as they allow unwanted interleavings to occur.
For example, if we simply substituted Memcached's blocking hash table and LRU by their non-blocking counterparts, unwanted interleavings would make it possible for cached items to be in present in one of the structures and absent in the other.
Maintaining correctness in such setting would require additional work, which in turn would negatively impact performance.
\systemname does not have a separate data structure to implement an eviction policy, but has it embedded into its hash table instead.
% \CLOCK
To this end, we took inspiration from \CLOCK, a known Operating Systems page replacement algorithm.
\CLOCK~\cite{corbator:multics-paging} associates a bit (which we refer to as the \CLOCK value) to each cache entry: the value~$0$ represents a not recently used item, while the value~$1$ represents a recently used item.
Cache eviction using a very fine-grained approach, based in associating a \CLOCK value to each item in \systemname, would require the eviction procedure to traverse all hash table buckets.
As each hash table bucket is implemented as a linked-list, with items that are not stored continuously in memory, each step in the list traversal would likely need to fetch data from main memory.
This means that traversals would not be cache effective and, thus, exhibit lower performance.
In \systemname we opted for a medium-grained approach, associating instead a \CLOCK value to each hash table bucket. This means that the eviction is cache friendly, as it traverses continuous blocks of memory.
Since the hash table expansion occurs when the number of cache items is~$1.5\kern0.25pt\times$ the number of hash table buckets, we know that each \CLOCK value represents at most~$1.5$ items.
Furthermore, \systemname's \CLOCK values are not limited to just one bit, so that its eviction policy can distinguish between mildly and highly popular items.
% Adding Non-Blocking
Through a hash table with embedded \CLOCK eviction, we can now substitute \systemname's hash table blocking linked-list buckets by non-blocking linked-lists.
To do so, we make use of Harris' pragmatic non-blocking linked-list algorithm~\cite{harris2001pragmatic}.
%Memory Reclamation
To be able to reclaim memory while utilizing a non-blocking concurrency control strategy, we need to employ a memory reclamation scheme.
\systemname's memory reclamation scheme is based on DEBRA~\cite{debra}, due to its flexibility and performance.
DEBRA is a general memory reclamation scheme that does not assume the overlying algorithm knows when it is out of memory, meaning the memory reclamation scheme progresses even when no memory needs to be reclaimed.
Since \systemname is a caching system, it \textit{must} know when it is out of memory.
For this reason, \systemname~deviates from DEBRA by only progressing the memory reclamation scheme when it is absolutely necessary, minimizing the amount of work required to reclaim memory.
%hash table expansion
Finally, having a blocking hash table expansion algorithm partially nullifies some of the progress condition advantages that come with having non-blocking concurrency control.
Thus, \systemname implements a non-blocking hash table expansion algorithm, opposed to Memcached's stop-the-world like expansion.

%Evaluation
\paragraph{\bfseries Evaluation.}
Our evaluation aims to clarify some characteristics of \systemname and its intermediate step, where Memcached eviction policy was substituted by a hash table with embedded \CLOCK eviction (\MEMCLOCK):
\begin{itemize}
    \item What is the impact on both performance and hit-ratio of having an approximated LRU eviction policy (\MEMCLOCK) rather than a strict one?
    \item How does each system perform under varying degrees of contention?
\end{itemize}

Contention was mediated by a few characteristics of the system, its environment and the workload it is being subjected to. In particular, the system's contention degree was dependent on the item size, item access frequency, and network bandwidth.

\begin{figure}[h]
    \centering
    \subfloat[Throughput.]{%
        \includegraphics[width=0.47\textwidth]{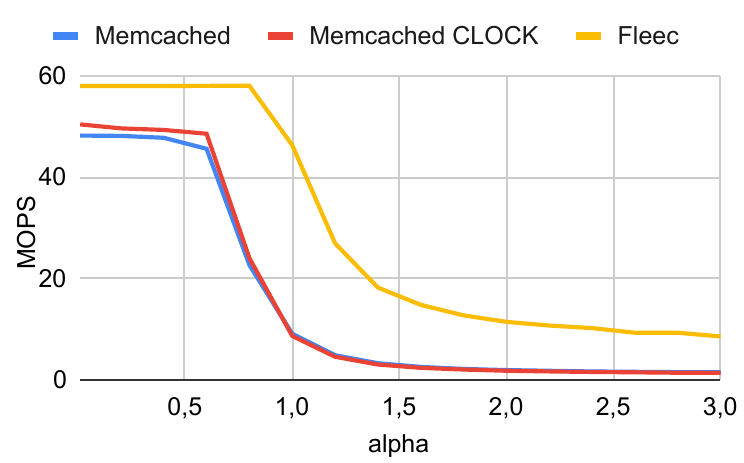}%
        \label{fig:alphastudy}%
    }
    \hfill
    \subfloat[Speedup.]{%
        \includegraphics[width=0.47\textwidth]{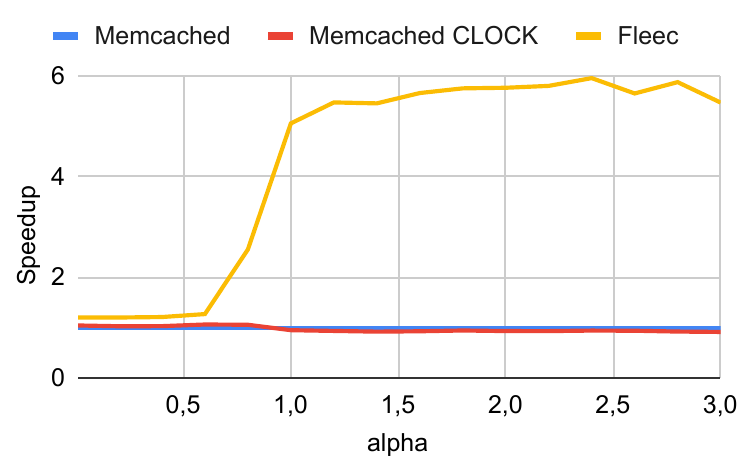}%
        \label{fig:alphastudyspeedup}%
    }
    \caption{Throughput and Speedup varying zipfian $\alpha$, for a read-intensive (99\% reads) workload.}
    \label{figs:tests}
\end{figure}

\autoref{figs:tests} depicts the throughput of Memcached, \MEMCLOCK, and \systemname, under read-intensive (99\% reads) workloads and varying degrees of skewness (the higher the $\alpha$ value, the more skewed the access distribution is).
These tests were performed with small items, so network communication is not a performance bottleneck.

From \autoref{fig:alphastudy}, it becomes evident that \systemname consistently provides higher throughput under every workload.
\autoref{fig:alphastudyspeedup} clarifies the relative performance between the three systems, emphasizing the impact of \systemname's new design and concurrency control strategy.

Our experimental results show that the \CLOCK-based eviction policy used in both \MEMCLOCK and \systemname does not  significantly impact the hit-ratio.
In terms of performance and latency, \MEMCLOCK exhibits throughput and latency similar to the original Memcached.
\systemname, on the other hand, can obtain up to~$6\kern0.25pt\times$ improvement in throughput and up to~$1/6$ of the latency, w.r.t.\ Memcached and  when under very high contention scenarios; $\approx\kern-3.25pt 1.2\kern0.25pt\times$ improvement under medium contention scenarios; and equivalent (to Memcached) performance under low contention scenarios.

%Conclusion
\paragraph{\bfseries Conclusion.}
\systemname performs significantly better than Memcached under high- and mid-contention scenarios, while presenting no performance degradation under low contention scenarios.  Thus, \systemname can easily be used as a plug-in replacement of Memcached, with no identified disadvantages.

%Acknowledgments
\paragraph{\bfseries Acknowledgments.}
This work was supported by NOVA LINCS (UIDB/04516/2020) with the financial support of FCT.IP.

%% file: bibliography.bib
@online{memcached,
    url = {http://memcached.org/},
    urldate = {2022-12-28},
    title = {Memcached},
}

@inproceedings{harris2001pragmatic,
  title={A pragmatic implementation of non-blocking linked-lists},
  author={Harris, Timothy L},
  booktitle={International Symposium on Distributed Computing},
  pages={300--314},
  year={2001},
  organization={Springer}
}

@inproceedings{debra,
  doi = {10.1145/2767386.2767436},
  url = {https://doi.org/10.1145/2767386.2767436},
  year = {2015},
  month = jul,
  publisher = {{ACM}},
  author = {Trevor Alexander Brown},
  title = {Reclaiming Memory for Lock-Free Data Structures},
  booktitle = {Proceedings of the 2015 {ACM} Symposium on Principles of Distributed Computing}
}

@TECHREPORT {corbator:multics-paging,
	ADDRESS = "Cambridge, MA" ,
	AUTHOR = "Corbato, F." ,
	INSTITUTION = "M.I.T." ,
	MONTH = "July" ,
	NUMBER = "MAC-M-384" ,
	TITLE = "A Paging Experiment with the {M}ultics System" ,
	TYPE = "Project MAC" ,
	YEAR = "1968" ,
	COMMENT = "Multics, Paging",
	LOCATION = "6.033 Notes"
}
